%% file: paper.tex
\documentclass[conference, a4paper]{IEEEtran}
\usepackage{cite}
\usepackage{amsmath,amssymb,amsfonts}
\usepackage{algorithmic}
\usepackage{graphicx}
\usepackage{textcomp}
\usepackage{xcolor}
\usepackage{csquotes}
\usepackage{subcaption}
\usepackage{booktabs}
\usepackage{multirow}
\usepackage{url}
\usepackage{siunitx}

\usepackage{siunitx}
\DeclareSIUnit{\belmilliwatt}{Bm}
\DeclareSIUnit{\dBm}{\deci\belmilliwatt}
\DeclareSIUnit{\dBi}{\deci\bel i}

\usepackage[nolist]{acronym} 

\usepackage{xcolor,colortbl}

\definecolor{DDGray}{gray}{0.8}
\definecolor{DGray}{gray}{0.82}
\definecolor{Gray}{gray}{0.85}
\definecolor{LGray}{gray}{0.9}
\definecolor{LLGray}{gray}{0.95}

\newcolumntype{e}{>{\columncolor{DDGray}}c}
\newcolumntype{f}{>{\columncolor{DGray}}c}
\newcolumntype{g}{>{\columncolor{Gray}}c}
\newcolumntype{h}{>{\columncolor{LGray}}c}
\newcolumntype{i}{>{\columncolor{LLGray}}c}

\usepackage[inline,shortlabels]{enumitem}

\begin{document}
\bstctlcite{IEEEexample:BSTcontrol}
\title{A5/1 is in the Air: Passive Detection of 2G (GSM) Ciphering Algorithms\\
}

\author{
	\IEEEauthorblockN{
		Matthias Koch\IEEEauthorrefmark{1},
		Christian Nettersheim\IEEEauthorrefmark{1}\IEEEauthorrefmark{3}, 
		Thorsten Horstmann\IEEEauthorrefmark{1}\IEEEauthorrefmark{3}, 
		Michael Rademacher\IEEEauthorrefmark{1}\IEEEauthorrefmark{3}  
	}
	\IEEEauthorblockA{\IEEEauthorrefmark{1}Fraunhofer FKIE, Bonn, Germany, {firstname.lastname}@fkie.fraunhofer.de}
    \IEEEauthorblockA{\IEEEauthorrefmark{3}Hochschule Bonn Rhein-Sieg, Sankt Augustin, Germany, {firstname.lastname}@h-brs.de}
}

\maketitle

\begin{abstract}
This paper investigates the ongoing use of the A5/1 ciphering algorithm within 2G GSM networks. Despite its known vulnerabilities and the gradual phasing out of GSM technology by some operators, GSM security remains relevant due to potential downgrade attacks from 4G/5G networks and its use in IoT applications.
We present a comprehensive overview of a historical weakness associated with the A5 family of cryptographic algorithms. Building on this, our main contribution is the design of a measurement approach using low-cost, off-the-shelf hardware to passively monitor Cipher Mode Command messages transmitted by base transceiver stations (BTS). We collected over 500,000 samples at 10 different locations, focusing on the three largest mobile network operators in Germany.
Our findings reveal significant variations in algorithm usage among these providers. One operator favors A5/3, while another surprisingly retains a high reliance on the compromised A5/1. The third provider shows a marked preference for A5/3 and A5/4, indicating a shift towards more secure ciphering algorithms in GSM networks.
\end{abstract}


\section{Introduction}\label{sec:introduction}
While some mobile operators have begun phasing out their \ac{GSM} networks, notably in the US~\cite{TMobileb13:online}, \ac{GSM} remains widely relevant globally due to its extensive coverage, which supports older cellular phones, \ac{IoT} devices like smart meters and \ac{GPS} trackers.
It also serves as a fallback option in areas where newer network generations do not provide sufficient coverage.
Moreover, although one might view 2G network security as outdated, it remains important to recognize that downgrade attacks from 4G/5G to 2G still pose a significant threat, making users vulnerable even when modern network generations are present~\cite{jcp4010002}.\\

Among the most prominent security issues is the vulnerability of the A5/1 algorithm.
Initially designed as a proprietary ciphering algorithm, A5/1's stream cipher procedure was reverse engineered in the late 1990s~\cite{A51PedImpl99:online}.
A5/1 is initialized by loading its three \acp{LFSR} with a 64-bit session key (called $K_c$) and a frame number as \ac{IV}. The resulting state is called the \emph{initial state} from which possible session keys can be extracted by back-clocking. %
Afterwards, the registers are clocked with a majority rule enabled for 100 additional rounds (the output of which is discarded) and then the subsequent key stream is used for the encryption of the next burst.
This led to a variety of research efforts attacking various mathematical weaknesses of its protocol.

While many of these early efforts were constrained by their dependency on large numbers of recorded messages or impractically long attack and pre-computation times, the first practical (near) real-time cracking attack of the session key was presented by Nohl et al. around 2010~\cite{nohl2010attacking}.
This attack leveraged several earlier insights.
For instance, well-known control messages, such as \texttt{\ac{CMC}} and \texttt{System Information} messages, can be exploited as known-plaintexts to recover the cipher stream sequence, largely due to the presence of numerous constant padding bits.
Combining distinguished points to reduce table lookup bottlenecks, substituting traditional \ac{TMTO} lookup tables with rainbow tables to minimize collisions as well as applying optimizing strategies for key space and table compression, the \emph{A5 Cracking Project}~\cite{nohl2010attacking, Lu2016} was able to make precomputed rainbow tables publicly available at a size of around \SI{2}{\tera\byte}.
These allow an attacker to efficiently recover candidates for the aforementioned \emph{initial state} from few known cipher streams yielding a session key with high probability.
A detailed report, highlighting the development and the interplay of these various techniques, can be found in \cite{Lu2016}.%

Open source cracking tools like \texttt{kraken}~\cite{github.kraken} combined with today's increased processing power and decreased costs for storage make these attack vectors available to the public - as can be seen by the existence of several YouTube tutorials that detail the whole setup and usage\footnote{Due to legal considerations, we do not include a reference to these videos.}. Possible mitigations exist in the form of padding randomization~\cite{nohl2010attacking, 3gpp.44.006} and a gradual move to the block-cipher-based A5/3 and A5/4 algorithms~\cite{nohl2010attacking}, but in the face of backward compatibility~\cite{3gpp.43.020}, implementation on both network and \ac{MS} side remains an open question.

Nowadays, in addition to A5/1, A5/3 and A5/4 are in use for encrypting over-the-air messages between a \ac{MS} and a \ac{BTS}. 
This raises the main research question of this work: \textbf{How much A5/1 is actually still used in today's mobile networks in Germany?}

The paper is organized as follows:
Section~\ref{sec:rel_work} presents related work regarding the main research question.
Section~\ref{sec:background} presents an overview of necessary \ac{GSM} network mechanics.
We designed our own measurement approach based on low-cost \ac{COTS} hardware which we introduce in Section~\ref{sec:methodology}.
Results are in Section~\ref{sec:evaluation}, limitations in Section~\ref{sec:limitations}, and Section~\ref{sec:conclusion} concludes the paper with future research directions.

\section{Related Work}
\label{sec:rel_work}
\subsection{A5 Usage Surveys}

Morgan \cite{morgan} proposed two methods for determining the usage and support of different A5 algorithms by Estonian mobile operators: a passive method and an active method.
The passive method aligns with our approach, utilizing a heuristic technique that involves counting \ac{CMC} messages transmitted by the \ac{BTS}.
These messages are received by \ac{SDR} devices, such as HackRF or RTL-SDR dongles, and processed using the \texttt{gr-gsm}\cite{github.grgsm} software.
Their active method involves connecting a modified phone to the network.
In one scenario, the phone is configured to support only one specific A5 algorithm, allowing for determination of whether a \ac{BTS} is capable of that algorithm.
In another scenario, the phone supports all available A5 algorithms, providing insight into which algorithm the \ac{BTS} prefers.
In the first scenario, a successful location update indicated support for the chosen A5 algorithm, while in the second, the algorithm was identified using an \ac{SDR}. The measurements focused on A5/1, A5/2, and A5/3, excluding A5/4.

The results showed that only one of three Estonian providers used A5/3, with the others exclusively using A5/1. Notably, none used A5/2, adhering to the recommendations of \cite{3gpp.43.020} and \cite{gsma_fs.35}.
Their passive method had limitations, including short measurement periods of 8-17 hours each and a small set of specific \ac{BTS} per provider, which prevented the observation of long-term effects.
In contrast, our work addressed these limitations by conducting long-term measurements across multiple locations, providing a more comprehensive view of the network's encryption activities.

\subsection{SnoopSnitch}
The SnoopSnitch app \cite{github.snoopsnitch} is another notable project that collects statistics on A5 algorithm usage as part of a broader mobile security analysis. Primarily designed to detect threats like IMSI catchers and silent/binary SMS, it also gathers data on mobile networks that users connect to. This data is compiled into yearly automated reports, which are published on the GSMmap website \cite{gsmmap.map}.
As noted in the SnoopSnitch FAQ \cite{snoopsnitch.faq}, the app has significant limitations. It requires a rooted Android device with a Qualcomm baseband chip to access raw radio messages via the \texttt{/dev/diag} interface. Compatibility is limited to Android versions 4.4 to 12, excluding other operating systems and certain custom ROMs or devices lacking the diagnostic interface kernel driver.

The 2023 GSMmap report for Germany \cite{gsmmap.report} indicates A5/1 usage between 13\% and 36\% across providers, though it notes that these figures are averages from diverse user contributions and \enquote{may be influenced by factors like location, \ac{SIM} type, and network load}\cite{snoopsnitch.faq}.
Moreover, as discussed in \cite{morgan}, SnoopSnitch only detects A5 algorithms used by test devices, not other users on the same network.
Our passive approach aims to address this limitation by collecting data on all devices within a cell, unlike methods that rely on individual user data.
In addition, GSMmap reports on the algorithms A5/0, A5/1, and A5/3.
Our approach cannot detect connections with A5/0, as the absence of encryption means that there are no \acp{CMC} that we passively eavesdrop on; however, we found many connections utilizing A5/4.

\section{Background on GSM}\label{sec:background}
To motivate our approach in Section~\ref{sec:methodology}, a brief overview of a typical connection setup in \ac{GSM} is provided hereafter.
The first access point of an \ac{MS} into the \ac{GSM} network via the radio interface is the \ac{BTS} of the respective cell. Each \ac{BTS} transmits information to the \ac{MS} on the downlink frequency (network $\rightarrow$ MS) and listens to incoming data from the \ac{MS} within its coverage area on the uplink frequencies (MS $\rightarrow$ network).
The frequencies that a \ac{BTS} can offer are selected from a subset of those available to the \ac{MNO}, ensuring that they do not interfere with frequencies used by neighboring cells.
The entire frequency range is typically licensed to various \acp{MNO} by national regulatory bodies, such as the \textit{Bundesnetzagentur} in Germany, for use in commercial \ac{GSM} operations.
In our context, we focus on the extended GSM-900 frequency bands (E-GSM) used in Germany, which utilize the downlink frequency range of \SIrange{925.0}{960.0}{\mega\hertz}.
Before the \ac{MS} can transmit and receive encrypted data, it must first establish a connection to the subscriber's network.
The exchange between the \ac{MS} and the network varies slightly depending on the \ac{MS}'s state prior to establishing a new connection, such as being idle, switched off, entering a new location area, or requesting an additional connection. The following steps are a condensed high-level overview ~\cite{sauter3, eberspaecher, heine, 3gpp.24.008}:
\begin{enumerate}[leftmargin=*]
    \item The \ac{BTS} periodically broadcasts \texttt{System Information} messages over the \ac{BCCH}, providing details about its associated \ac{MNO}, location area, and cell identity.
    \item The \ac{MS} measures the signal strength of the available frequencies from the GSM range, whereby the \acp{BCCH} are monitored.
    \item The \ac{MS} selects the \ac{BTS} with the strongest signal matching the provider identity stored in its \ac{SIM} card. 
    \item A signalling channel is established, allowing the \ac{MS} to request a service, such as a voice call or SMS.
    \item To inform the network of its current location, the \ac{MS} initiates a location update via a \texttt{Location Update Request}. This request includes various identifiers for the \ac{MS} and a \texttt{Mobile Station Classmark} message, which notifies the network of the supported A5 algorithms.
    \item The \ac{MS} and network perform identification and authentication procedures. During this process, a new session key $K_c$ for ciphering is derived on both ends using a 128-bit random challenge \texttt{RAND}, chosen by the network, and the master key $K_i$ stored on the \ac{SIM} card and at the network's \ac{AuC}, respectively.
    \item The network selects a ciphering algorithm supported by both the \ac{MS} and the network for the connection and sends a \texttt{\acf{CMC}} message, informing the \ac{MS} of its choice, to initiate encrypted communication.
    \item This command causes the \ac{MS} to enable (de-)ciphering for all subsequent messages of the connection and responds with an encrypted \texttt{Cipher Mode Complete}.
    \item If the network deciphers the \texttt{Cipher Mode Complete} message, the location update completes and service requests proceed over an encrypted connection.
\end{enumerate}

\section{Methodology}\label{sec:methodology}
As seen in Section \ref{sec:background}, one approach to gathering data on the types of encryption algorithms used by a provider is to listen passively to the \texttt{\acfp{CMC}} sent by a \ac{BTS} on the downlink.

\begin{figure}[ht]
    \centering
    \includegraphics[width=\columnwidth]{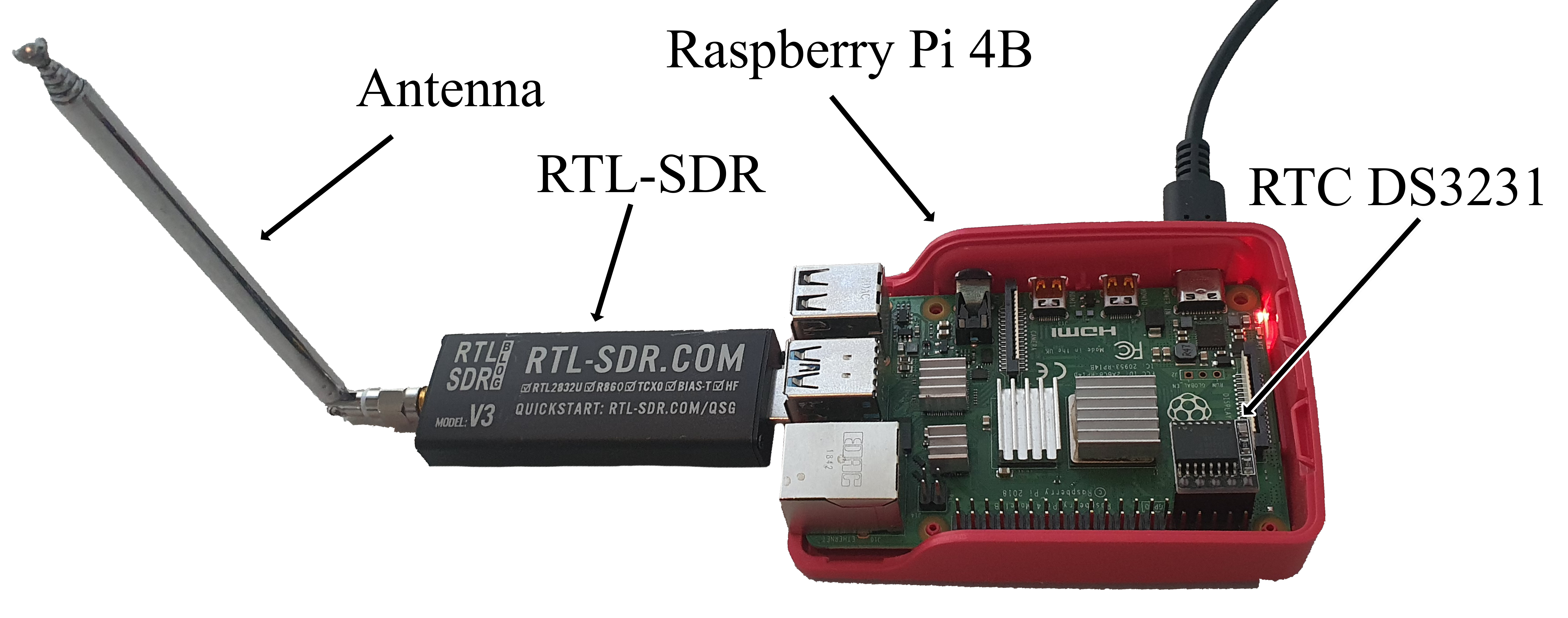}
    \caption{Components of our sensor}
    \label{fig:sensor}
\end{figure}

\subsection{Sensor}\label{sec:sensor}
To capture and log these \acp{CMC} over long periods of time, we utilize an \ac{SDR} to listen to downlink traffic of a \ac{BTS} and simultaneously filter out instances of \acp{CMC} from the incoming packets.
For each captured \ac{CMC}, we extract the cipher algorithm used and log the timestamp of transmission.
Our sensors, as depicted in Figure \ref{fig:sensor}, consist of an RTL-SDR dongle connected to either a monopole or dipole antenna and a Raspberry Pi 4 Model B. %
The data collection is controlled by the \texttt{gsm-monitor.service}~\cite{GSMCryptoSniffingRepo}, a custom \texttt{systemd} service. This service does the following:
\begin{enumerate}[leftmargin=*]
    \item To start out, it searches for all available \ac{GSM} frequencies received by the dongle. This uses the \texttt{kalibrate\_rtl}\cite{github.kalibrate_rtl} tool to find a list of frequencies and their respective signal strength.
    \item The \texttt{gsm-monitor.service} is configured with a provider's \ac{MNC}.
    The frequencies found in the first step are then sorted by signal strength and the according channel frequencies are probed for a few seconds to find the \ac{MNC} from the \texttt{\ac{SI3}}.
    To receive \ac{GSM} packets with the \ac{SDR}, we use \texttt{gr-gsm}\cite{github.grgsm}.
    When a frequency for the desired provider is found, the \ac{LAC} and the \ac{CID} are additionally recorded.
    This effectively finds the strongest currently available frequency for the chosen provider.
    \item %
    The data points each consist of a timestamp and an algorithm identifier, see \cite{3gpp.44.018}.
    Filtering and extracting is done with \texttt{tshark}\cite{tshark}.
    \item To ensure continuous sufficient signal strength for data capture, a watchdog runs every 5 minutes, counting \ac{SI3} messages over 30 seconds, and restarts the service if the count falls below a set threshold.
\end{enumerate}
For further technical details and setup instructions, refer to the project's git repository~\footnote{\url{https://github.com/mclab-hbrs/GSM-Cipher-Sensor}}.

\subsection{Deployment}
We evaluated two distinct deployment options: one offline and the other online. %
For offline deployment, a real-time clock such as a \textit{DS3231} needs to be installed on the Raspberry Pi. %
For online deployments, time synchronization can be achieved using the \ac{NTP}.

Online deployments offer greater flexibility, but necessitate an active internet connection.
This connection can be preconfigured using the NetworkManager service before deployment.
To enhance accessibility, we opted for remote management of the monitoring stations via Tailscale, which provides \texttt{ssh} access from virtually any network \cite{tailscale}.
This setup allows us to manually administer changes to the chosen provider and retrieve collected data at any time.

\subsection{Measuring campaign}\label{sec:measuring-campaign}

\begin{table}[ht]
\centering
\begin{tabular}{r|rr|rr|rr}
\multicolumn{1}{l|}{} &
  \multicolumn{2}{c|}{\textbf{Provider A}} &
  \multicolumn{2}{c|}{\textbf{Provider B}} &
  \multicolumn{2}{c}{\textbf{Provider C}} \\
\multicolumn{1}{l|}{\textbf{Location}} &
  \multicolumn{1}{l}{\textbf{Days}} &
  \multicolumn{1}{l|}{\textbf{\#CMCs}} &
  \multicolumn{1}{l}{\textbf{Days}} &
  \multicolumn{1}{l|}{\textbf{\#CMCs}} &
  \multicolumn{1}{l}{\textbf{Days}} &
  \multicolumn{1}{l}{\textbf{\#CMCs}} \\ \hline
1/u  & 2.67  & 418   & 0     & 0      & 3.02  & 6131  \\
2/u  & 3.8   & 872   & 3.75  & 5669   & 3.77  & 8576  \\
3/u  & 58.08 & 13852 & 42.95 & 212567 & 15.07 & 40438 \\
4/s  & 35.76 & 34302 & 33.87 & 54477  & 18.29 & 54914 \\
5/s  & 1.07  & 966   & 1.06  & 1945   & 1.06  & 2914  \\
6/s  & 2.05  & 964   & 1.99  & 2238   & 3.45  & 9611  \\
7/u  & 7.15  & 1756  & 7.46  & 11780  & 1.99  & 7454  \\
8/r  & 4.75  & 1778  & 2.87  & 3999   & 2.87  & 50247 \\
9/s  & 1.42  & 946   & 2.23  & 5186   & 2.27  & 9742  \\
10/r & 2.13  & 1189  & 3.72  & 20182  & 0     & 0    
\end{tabular}
\caption{Overview of the measurement campaign. Duration of the measurements and the number of captured \acp{CMC} per provider. Settlement type of the location given by u~=~urban, s~=~suburban, r~=~rural.}
\label{tab:campaign_overview}
\end{table}

In total, we constructed five sensors as outlined in Section~\ref{sec:sensor}. %
A total of 565,115 \acp{CMC} packets were analyzed at 10 distinct locations.
Measurements were conducted over a period of 88 days, commencing on 23 December 2024 and concluding on 19 March 2025.
Furthermore, the measurement duration differs at each location, as seen in Table \ref{tab:campaign_overview}.
The locations at which we measured were distributed in and around Bonn, Germany.
There were 10 locations in a total of 6 distinct municipalities.
An assessment of whether the locations are urban, suburban or rural is also given in Table \ref{tab:campaign_overview}.
Altogether, we measured in four urban, four suburban and two rural locations.
The results later showed that location did not have a major influence on the outcomes.

At least one day was measured per location and per provider, but usually more.
As we were limited by the number of sensors, in most locations the provider to be measured was replaced after a period of time, so that the providers were not measured at the same time, but one after the other.

Long-term measurements were carried out at locations 3 and 4, which lasted approx. 73 days and approx. 54 days respectively.
This was intended to generate a good initial pool of measured values.
No measurements were taken for Provider B at location 1 and Provider C at location 10.
At location 1, the signal for Provider B experienced excessive attenuation, presumably due to the indoor measurement environment, resulting in an insufficient signal strength for a valid measurement to be obtained.
At Location 10, Provider C could also not be measured due to insufficient signal strength.

\section{Evaluation \& Results}\label{sec:evaluation}

In this section, we will present and discuss the results.
Our results are examined across the three largest providers in Germany which are pseudonymized as Provider A, Provider B and Provider C.

Because we took measurements at the locations for different durations, the absolute values were not included in the overall result.
Otherwise, the results would be skewed by the long-term measurements from locations 3 and 4.
As we cannot rule out the possibility of site-specific anomalies at the locations, we weighted the measurement. For example, we measured all (10) locations for Provider A, but only 9 locations for provider B and C. Therefore, for the overall results, Provider A has weighting of $\frac{1}{10}$ and Provider B and C of $\frac{1}{9}$ respectively.\\

\begin{figure}[ht]
    \centering
    \includegraphics[width=\columnwidth]{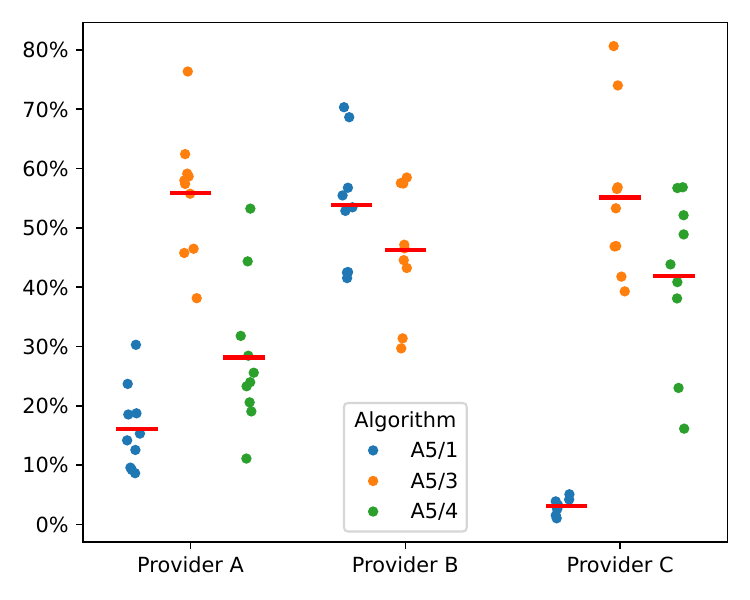}
    \caption{Distribution of algorithm usage for the different providers across the various locations. Mean values are marked with red line.}
    \label{fig:stripplot}
\end{figure}
In the analysis presented in Figure~\ref{fig:stripplot}, we examine the distribution of algorithm usage (A5/1, A5/3, A5/4) across the providers. Each point represents the usage percentage of an algorithm at a specific location, while the red horizontal lines indicate the mean usage for each algorithm per provider.

\begin{itemize}[leftmargin=*]
    \item The result for \textbf{Provider A} shows a strong use of A5/3 with an average of 55.8\%, whereby A5/4 is at around 28.1\%.
    Algorithm A5/1 is still frequently used with approx. 16.1\% on average.
    \item With \textbf{Provider B} it is even 53.8\% for algorithm A5/1, whereas A5/3 is at 46.2\%.
    We were unable to detect any communication for algorithm A5/4 at Provider B.
    We cannot say whether our observations occurred by chance, but it is statistically reasonable to assume that Provider B does not support A5/4 and even favors A5/1.
    \item \textbf{Provider C} has the lowest proportion of A5/1 traffic at around 3\%. The remaining share is distributed between A5/3 with 55.1\% and A5/4 with 41.8\%.
\end{itemize}

As discussed in Section~\ref{sec:introduction}, algorithm A5/1 is deemed broken. %
That is reflected in its low utilization rate at both Provider C and Provider A.
In contrast, the usage at Provider B is surprisingly the highest, which raises questions about its operational context and the reasons behind this choice, despite the known issues associated with this algorithm.
The complete absence of A5/4 at Provider B also raises concerns about the security standards of the network.

One theory is that since \ac{MS} and \ac{BTS} negotiate the algorithm and older \ac{MS} tend to choose the older algorithm. Newer equipment would also prefer newer mobile technologies such as 4G or 5G. Another possibility is that the \ac{BTS} does not implement, for example, the A5/4 algorithm. With our passive approach, we are not able to further test this.

\begin{figure}[ht]
    \centering
    \includegraphics[width=\columnwidth]{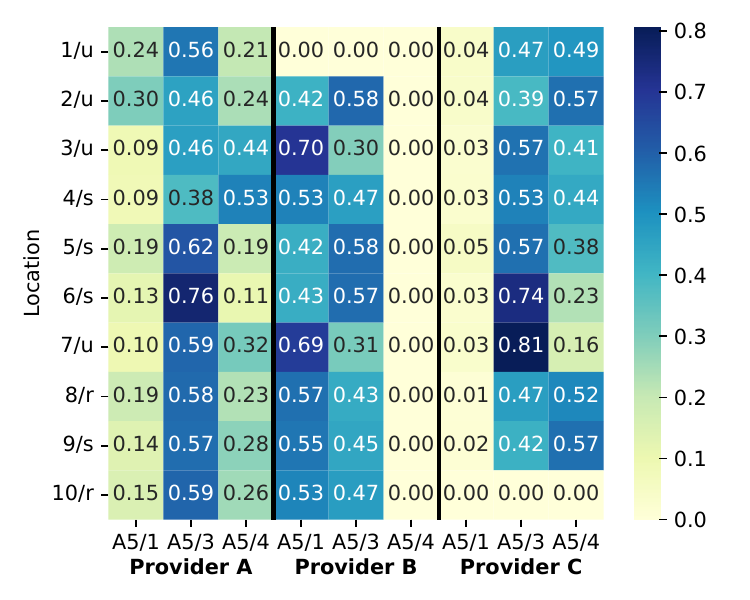}
    \caption{Heatmap showing the usage rates of the algorithms per provider and per location. The color intensity represents the rate of algorithm usage. Settlement type of the location given by u~=~urban, s~=~suburban, r~=~rural.}
    \label{fig:heatmap}
\end{figure}

Figure \ref{fig:heatmap} shows a heatmap indicating the proportions of the algorithms for each location and provider.
Darker fields indicate a more frequent occurrence than lighter fields.
As already mentioned in Section~\ref{sec:measuring-campaign} we would like to point out the missing measurements from Provider B in Location 1 and Provider C in Location 10.

As the stripplot in Figure \ref{fig:stripplot} has already shown, it is especially the high utilization of A5/1 and the lack of A5/4 at Provider B that stands out.
A particularly large amount of A5/1 was measured during the long-term measurement in location 3.
This may be due to a few \acp{MS} negotiating this algorithm.
With our method, it is not possible to make more detailed statements about this.
Usage varies significantly between locations, but the core findings hold.
The consistently low usage of A5/1 at Provider C is also noteworthy.

\begin{figure}[ht]
    \centering
    \includegraphics[width=\columnwidth]{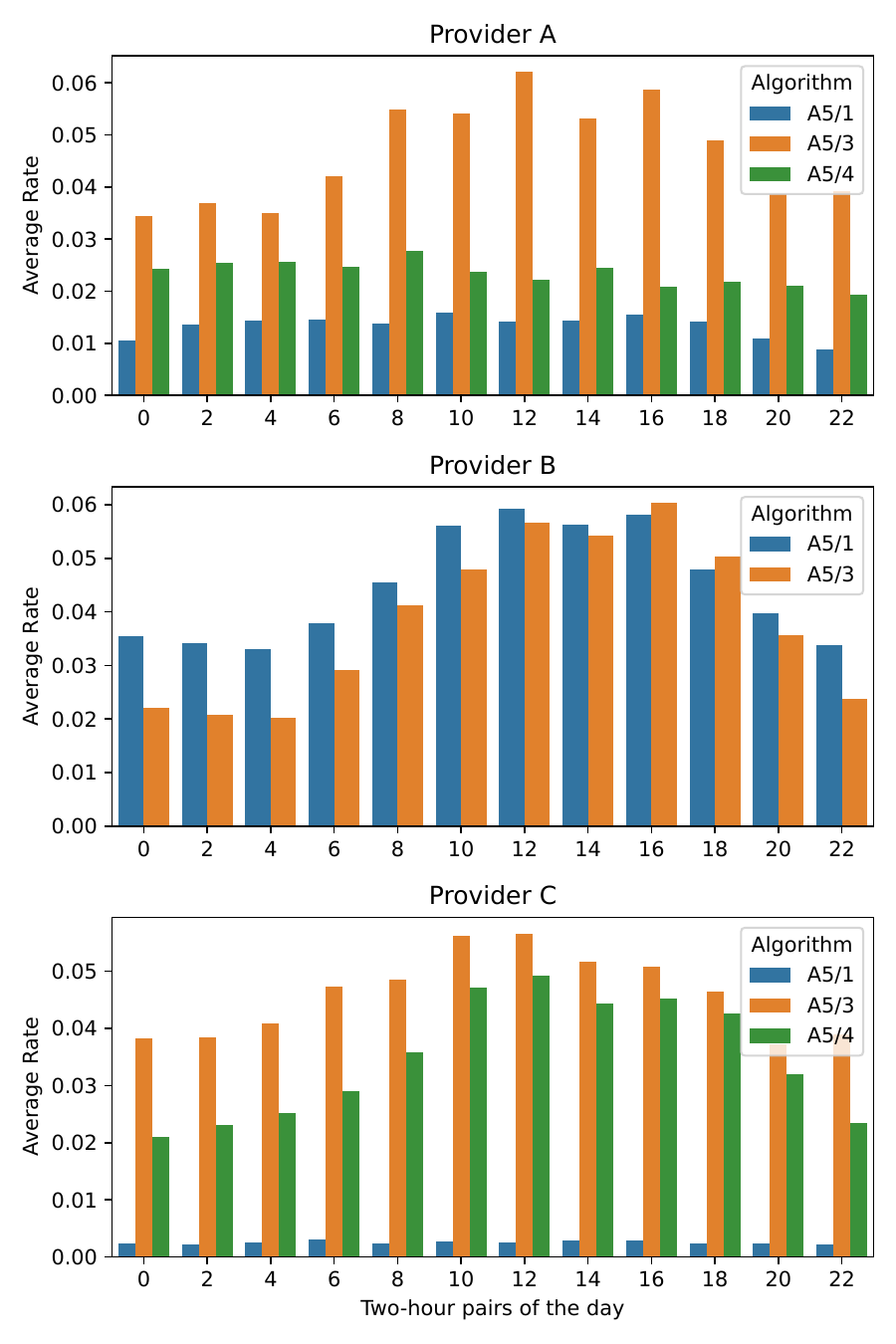}
    \caption{Normalized hourly usage rates of different encryption algorithms for each provider. The rates are calculated as the proportion of the total measurements for each location and hour. The graph shows the average rate across all locations for each hour of the day.}
    \label{fig:alg_usage_hourly}
\end{figure}

Figure~\ref{fig:alg_usage_hourly} shows the algorithm usage by time of day.
In this graph, each location contributes 1/9 or respectively 1/10 to the graph.
Two-hour pairs are listed on the x-axis, i.e. 0 represents hours 0-1, 2 represents hours 2-3, etc.
The values of the graph, i.e. the y-axis, were formed as follows:
First, for each location we calculated how much of the total traffic each algorithm has at each hour.
Then the average was taken for each algorithm at each hour across the locations.
This means that for each diagram all the bars add up to 1.
By doing so we preserve the distribution of usage throughout the day.

\begin{itemize}[leftmargin=*]
    \item At \textbf{Provider B}, the use of the two algorithms is similar to each other over the course of the day.
    The graph corresponds roughly to what you would expect if you assume that the traffic is largely produced by humans or machines operated by humans.
    Comparatively little traffic in the nighttime hours, slowly increasing at the peak traffic times of midday and afternoon and then decreasing again towards the evening and at night.
    \item The graph for \textbf{Provider C} follows a similar pattern to that of Provider B for algorithms A5/3 and A5/4.
    For algorithm A5/1, the traffic is at a relatively constant low level.
    This could be a sign that the traffic does not come from humans, but from machines such as sensors or actuators.
    These machines may be older, meaning that the newer algorithms could be not supported.
    \item By contrast, only algorithm A5/3 follows the pattern for \textbf{Provider A}.
    Algorithms A5/1 and A5/4 behave rather consistently over the course of the day, with some upward and downward fluctuations.
    While the behavior of algorithm A5/1 at provider C could still be explained by machine traffic, this explanation is more difficult here. 
\end{itemize}

Our evaluation of algorithm usage across the three major providers in Germany reveals significant variations in the adoption of encryption algorithms.
Provider A demonstrates a notable preference for A5/3, while Provider B's unexpectedly high usage of the broken A5/1 raises concerns about its security practices.
Provider C, on the other hand, displays a predominantly low utilization of A5/1, indicating a shift towards more secure alternatives like A5/3 and A5/4.

\section{Known limitations}\label{sec:limitations}
By design, our approach is limited to monitoring packets transmitted over the air when communicating with unknown parties.
We rely on the traffic generated between \ac{MS} and \ac{BTS}, as our passive monitoring method can only capture what is already in the air.
However, by conducting long-term measurements across multiple locations, we can provide strong indicators of the cipher algorithm usage patterns within the monitored region.

Our current sensor setup has limitations due to the use of a single-frequency \ac{SDR}, which prevents simultaneous monitoring of multiple providers with a single sensor. 
Additionally, \ac{GSM}-1800 networks could be investigated with the present method, but as the deployed RTL-SDR dongles only receive frequencies up to \SI{1.75}{\giga\hertz} and as German \acp{MNO} gradually move their 2G operations into the \ac{GSM}-900 band to free up frequencies for LTE usage, we decided to limit both the scope of the survey and hardware cost.
These points could be addressed by either deploying multiple sensors or using more advanced \acp{SDR} capable of handling multiple frequencies and a higher frequency range in exchange for higher deployment costs.
The setup also requires a nearby power supply and indoor placement, limiting deployment options.
While battery-assisted, weatherproof designs could expand placement possibilities, this may not be necessary in areas sufficiently covered by 2G and densely populated like Germany.

Power consumption is another consideration.
The Raspberry Pi's continuous processing of \ac{GSM} packets creates a constant CPU load, which could quickly drain a battery.
This makes it challenging to use in areas without easy access to power supplies and likely precludes the use of more power-efficient IoT devices, as they typically rely on idle times or low-power operation periods that are not adequate in this application.

The limited CPU and USB bus speed of the Raspberry Pi may occasionally lead to randomly dropped \ac{GSM} packets.
However, since \acp{CMC} are sparsely distributed, compared to the rest of the packets, this doesn't significantly affect the overall observed patterns in cipher algorithm distribution over extended periods.
Computers with higher CPU and bus speeds could be used to avoid dropping, but in exchange for higher costs and less deployment flexibility.
Factors such as receiver placement, signal strength, and \ac{SDR} hardware quality may also influence data gathering.
Despite these limitations, the method provides valuable insights into long-term trends in cipher algorithm usage.

\section{Conclusion \& Open Questions}\label{sec:conclusion}
In conclusion, this study provides critical insights into the ongoing use of the A5/1 ciphering algorithm within 2G (\ac{GSM}) networks, despite its well-documented vulnerabilities.
Our findings reveal a significant variance in the adoption of cryptographic algorithms among major mobile network operators in Germany.
Notably, Provider B's high reliance on the compromised A5/1 algorithm raises serious concerns regarding its security practices, especially when juxtaposed with Providers A and C, which demonstrate a transition towards more secure alternatives like A5/3 and A5/4.
The persistence of A5/1 highlights the challenges posed by legacy systems and the potential risks of downgrade attacks as 4G/5G users switch between different generations of mobile networks.
Our methodology, employing low-cost hardware for passive monitoring, effectively captures the algorithm usage patterns, underscoring the importance of continued vigilance and assessment of network security practices.

Future research should focus on investigating the underlying reasons for the discrepancies in algorithm adoption among operators, the implications for user security, and the potential for upgrading legacy systems.
Additionally, there is a need to explore the feasibility of implementing more robust encryption standards across all operators to mitigate risks associated with outdated protocols.

Open questions remain regarding the long-term impacts of maintaining outdated encryption standards in a rapidly evolving technological landscape and how best to balance legacy support with enhanced security measures for users.

\input{acro}

\bibliographystyle{IEEEtran}
\bibliography{./bibabbrv.bib, ./lit.bib}
\end{document}

%% file: acro.tex
\begin{acronym}[]
\acro{3GPP}{3rd Generation Partnership Project}
\acro{AC}{Access Category}
\acro{ACK}{acknowledgement}
\acro{AI}{Abstract Interface}
\acro{AIFS}{Arbitration Interframe Space}
\acro{AODV}{Ad hoc On-Demand Distance Vector Routing Protocol}
\acro{AP}{Access Point}
\acro{API}{application programming interface}
\acro{APN}{Access Point Name}
\acro{ARP}{Address Resolution Protocol}
\acro{ARQ}{Automatic Repeat reQuest}
\acro{AS}{Autonomous System}
\acro{ASCII}{American Standard Code for Information Interchange}
\acro{ATIS}{Alliance for Telecommunications Industry Solutions}
\acro{ATM}{Asynchronous Transfer Mode}
\acro{AMR}{Automatic Meter Reading}
\acro{B.A.T.M.A.N.}{Better Approach to MANET}
\acro{BER}{Bit Error Rate}
\acro{BFS-CA}{Breadth First Search Channel Assignment}
\acro{BGP}{Border Gateway Protocol}
\acro{BPSK}{Binary Phase-Shift Keying}
\acro{BRA}{Bidrectional Routing Abstraction}
\acro{BS}{Base Station}
\acro{BSI}{Bundesamt für Sicherheit in der Informationstechnik}
\acro{BSSID}{Basic Service Set Identification}
\acro{BTS}{Base Transceiver Station}
\acro{CA}{Channel Assignment}
\acro{CAPEX}{capital expenditure}
\acro{CAPWAP}{Control And Provisioning of Wireless Access Points}
\acro{CARD}{Channel Assignment with Route Discovery}
\acro{CAS}{Channel Assignment Server}
\acro{CCA}{Clear Channel Assessment}
\acro{CDMA}{Code Division Multiple Access}
\acro{CF}{CompactFlash}
\acro{CIDR}{Classless Inter-Domain Routing}
\acro{CLICA}{Connected Low Interference Channel Assignment}
\acro{CLI}{Command Line Interface}
\acro{COTS}{Commercial Off-the-Shelf}
\acro{CPE}{Customer Premises Equipment}
\acro{CPU}{Central Processing Unit}
\acro{CRAHN}{Cognitive Radio Ad-Hoc Network}
\acro{CRCN}{Cognitive Radio Cellular Network}
\acro{CR}{Cognitive Radio}
\acro{CR-LDP}{Constraint-based Routing Label Distribution Protocol}
\acro{CRN}{Cognitive Radio Network}
\acro{CRSN}{Cognitive Radio Sensor Network}
\acro{CRVN}{Cognitive Radio Vehicular Network}
\acro{CSMA/CA}{Carrier Sense Multiple Access/Collision Avoidance}
\acro{CSMA}{Carrier Sense Multiple Access}
\acro{CSMA/CD}{Carrier Sense Multiple Access/Collision Detection}
\acro{CSV}{Comma-Separated Values}
\acro{CTA}{Centralized Tabu-based Algorithm}
\acro{CW}{Contention Window}
\acro{CWLAN}{Cognitive Wireless Local Area Network}
\acro{CWMN}{Cognitive Wireless Mesh Network}
\acro{DAD}{Duplicate Address Detection}
\acro{DCF}{Distributed Coordination Function}
\acro{DCiE}{Data Center Infrastructure Efficiency}
\acro{DDS}{Direct digital synthesizer}
\acro{DFS}{Dynamic Frequency Selection}
\acro{DGA}{Distributed Greedy Algorithm}
\acro{DHCP}{Dynamic Host Configuration Protocol}
\acro{DIFS}{Distributed Interframe Space}
\acro{DMesh}{Directional Mesh}
\acro{D-MICA}{Distributed Minimum Interference Channel Assignment}
\acro{DR}{Designated Router}
\acro{DSA}{Dynamic Spectrum Allocation}
\acro{DSLAM}{Digital Subscriber Line Access Multiplexer}
\acro{DSL}{Digital Subscriber Line}
\acro{DSR}{Dynamic Source Routing Protocol}
\acro{DSSS}{Direct-Sequence Spread Spectrum}
\acro{DTCP}{Dynamic Tunnel Configuration Protocol}
\acro{DVB}{Digital Video Broadcast}
\acro{DVB-H}{Digital Video Broadcast - Handheld}
\acro{DVB-RCS}{Digital Video Broadcast - Return Channel Satellite}
\acro{DVB-S2}{Digital Video Broadcast - Satellite - Second Generation}
\acro{DVB-S}{Digital Video Broadcast - Satellite}
\acro{DVB-SH}{Digital Video Broadcast - Satellite services to Handhelds}
\acro{DVB-T2}{Digital Video Broadcast - Second Generation Terrestrial}
\acro{DVB-T}{Digital Video Broadcast - Terrestrial}
\acro{E2CARA-TD}{Energy Efficient Channel Assignment and Routing Algorithm – Traffic Demands}
\acro{ECN}{Explicit Congestion Notification}
\acro{ECDF}{Empirical Cumulative Distribution Function}
\acro{EDCA}{Enhanced Distributed Coordination Access}
\acro{EDCF}{Enhanced Distributed Coordination Function}
\acro{EGP}{Exterior Gateway Protocol}
\acro{EICA}{External Interference-Aware Channel Assignment}
\acro{EIFS}{Extended Interframe Space}
\acro{EIGRP}{Enhanced Interior Gateway Routing Protocol}
\acro{EIRP}{Equivalent Isotropically Radiated Power}
\acro{EPI}{energy proportionality index}
\acro{ERO}{Explicit Route Object}
\acro{ETSI}{European Telecommunications Standards Institute}
\acro{ETT}{Expected Transmission Time}
\acro{ETX}{Expected Transmission Counts}
\acro{EUI}{Extended Unique Identifier}
\acro{FCC}{Federal Communications Commission}
\acro{FCS}{Frame Check Sequence}
\acro{FDD}{Frequency Division Duplex}
\acro{FDMA}{Frequency Division Multiple Access}
\acro{FEC}{Forward Error Correction}
\acro{FIFO}{First-In-First-Out}
\acro{FLOPS}{Floating-Point Operations Per Second}
\acro{FRR}{Fast Reroute}
\acro{FSL}{Free-Space Loss}
\acro{FSPL}{Free-Space Path Loss}
\acro{GAN}{Generic Access Network}
\acro{GDP}{Gross Domestic Product}
\acro{GEO}{Geosynchronous Earth Orbit}
\acro{GMPLS}{Generalized Multiprotocol Label Switching}
\acro{GNSS}{Global Navigation Satellite System}
\acro{GPS}{Global Positioning System}
\acro{GRE}{Generic Routing Encapsulation}
\acro{GSE}{Generic Stream Encapsulation}
\acro{GSM}{Global System for Mobile Communications}
\acro{GW}{Gateway}
\acro{HAP}{High-Altitude Platform}
\acro{HCCA}{HCF controlled channel access}
\acro{HCF}{Hybrid Coordination Function}
\acro{HLR}{Home Location Register}
\acro{HOL}{Head-of-line}
\acro{HOLSR}{Hieracical Optimised Link State Routing}
\acro{HPC}{hardware performance counters}
\acro{HSLS}{Hazy-Sighted Link State Routing Protocol}
\acro{HWMP}{Hybrid Wireless Mesh Protocol}
\acro{IAX2}{Inter-Asterisk eXchange Version 2}
\acro{IBSS}{Independent Basic Service Set}
\acro{ICMP}{Internet Control Message Protocol}
\acro{ICT}{Information and Communication Technologie}
\acro{IEEE}{Institute of Electrical and Electronics Engineers}
\acro{IE}{Information Element}
\acro{IETF}{Internet Engineering Task Force}
\acro{IETF}{The Internet Engineering Task Force}
\acro{IFS}{Interframe Space}
\acro{ITU}{International Telecommunication Union}
\acro{ITU-R}{\ac{ITU} Radiocommunication Sector}
\acro{IGP}{Interior Gateway Protocol}
\acro{IGRP}{Interior Gateway Routing Protocol}
\acro{ILP}{Integer Linear Programming}
\acro{ILS}{Iterated Local Search}
\acro{IPFIX}{IP Flow Information Export}
\acro{IP}{Internet Protocol}
\acro{IPv4}{Internet Protocol}
\acro{IPv6}{Internet Protocol, Version 6}
\acro{ISI}{Inter-symbol interference}
\acro{IS-IS}{Intermediate system to intermediate system}
\acro{ISM}{Industrial, Scientific and Medical}
\acro{ISP}{Internet Service Provider}
\acro{JSON}{JavaScript Object Notation}
\acro{KPI}{Key-Performance-Indicator}
\acro{LAA}{Licensed-Assisted Access}
\acro{LDC}{Least Developed Countries}
\acro{LA-CA}{Load-Aware Channel Assignment}
\acro{LCOS}{LANCOM Operating System}
\acro{LDP}{Label Distribution Protocol}
\acro{Ld}{Log-distance}
\acro{LDPL}{Log-distance path loss}
\acro{LEO}{Low Earth Orbit}
\acro{LER}{Label Edge Router}
\acro{LGI}{Long Guard Interval}
\acro{LLTM}{Link Layer Tunneling Mechanism}
\acro{LMA}{Local Mobility Anchor}
\acro{LMP}{Link Management Protocol}
\acro{LoS}{Line of Sight}
\acro{LOS}{Line of Sight}
\acro{LQF}{Longest-Queue-First}
\acro{LS}{Link State}
\acro{LSP}{Label-Switched Path}
\acro{LSR}{Label-Switched Router}
\acro{LST}{Link-State-Table}
\acro{LTE}{Long Term Evolution}
\acro{LTE-M}{\ac{LTE} Machine Type Communication}
\acro{LWAPP}{Lightweight Access Point Protocol}
\acro{MAC}{Media Access Control}
\acro{MAG}{Mobile Access Gateway}
\acro{MANET}{Mobile Adhoc Network}
\acro{MBMS}{Multimedia Broadcast Multicast Service}
\acro{MCG}{multi-conflict graph}
\acro{MCI-CA}{Matroid Cardinality Intersection Channel Assignment}
\acro{MCS}{Modulation and Coding Scheme}
\acro{MDR}{MANET Designated Router}
\acro{MEO}{Medium Earth Orbit}
\acro{MICS}{Media Independent Command Service}
\acro{MIES}{Media Independent Event Service}
\acro{MIHF}{Media Independent Handover Function}
\acro{MIHF++}{Media Independent Handover Function++}
\acro{MIH}{Media Independent Handover}
\acro{MIIS}{Media Independent Information Service}
\acro{MILP}{Mixed Integer Linear Programming}
\acro{MIMF}{Media Independent Messaging Function}
\acro{MIMO}{Multiple Input Multiple Output}
\acro{MNO}{Mobile Network Operator}
\acro{MIMS}{Media Independent Messaging Service}
\acro{MIPS}{Million Instruction Per Second}
\acro{MMF}{Mobility Management Function}
\acro{MMS}{Manufacturing Message Specification}
\acro{MN}{Mesh Node}
\acro{MonF}{Monitoring Function}
\acro{MPDU}{MAC Protocol Data Unit}
\acro{MPEG}{Moving Picture Experts Group}
\acro{MPE}{Multi Protocol Encapsulation}
\acro{MPLCG}{Multi-Point Link Conflict Graph}
\acro{MPLS}{Multiprotocol Label Switching}
\acro{MPLS}{Multi Protocol Label Switching}
\acro{MPLS-TE}{Multi Protocol Label Switching - Traffic Engineering}
\acro{MP}{Merge Point}
\acro{MPR}{Multipoint Relay}
\acro{MR-MC WMN}{Multi-Radio Multi-Channel Wireless Mesh Network}
\acro{MSC}{Mobile-services Switching Centre}
\acro{MSDU}{MAC Service Data Unit}
\acro{MSTP}{Mobility Services Transport Protocol}
\acro{MT}{Mobile Terminal}
\acro{MTU}{Maximum Transmission Unit}
\acro{MQTT}{MQ Telemetry Transport}
\acro{NAV}{Network Allocation Vector}
\acro{ns-3}{network simulator 3}
\acro{NB-IoT}{Narrowband \ac{IoT}}
\acro{NBMA}{Non-broadcast Multiple Access}
\acro{NetEMU}{Network Emulator}
\acro{NIDD}{Non-\ac{IP} Data Delivery}
\acro{NLOS}{None Line of Sight}
\acro{NMEA}{National Marine Electronics Association}
\acro{NPC}{Normalized Power Consumption}
\acro{NP}{Nondeterministic Polynomial Time}
\acro{NSIS}{Next Steps in Signaling}
\acro{NTP}{Network Time Protocol}
\acro{OFDMA}{Orthogonal Frequency Division Multiple Access}
\acro{OFDM}{Orthogonal Frequency Division Multiplex}
\acro{OLSR}{Optimized Link State Routing}
\acro{OPEX}{operational expenditure}
\acro{OSA}{Opportunistic Spectrum Access}
\acro{OSI}{Open Systems Interconnection}
\acro{OSPF}{Open Shortest Path First}
\acro{O2I}{Outdoor-to-Indoor}
\acro{OSPF-TE}{Open Shortest Path First - Traffic Engineering}
\acro{OVS}{Open vSwitch}
\acro{P2MP}{Point To Multipoint}
\acro{P2P}{Point To Point}
\acro{PA}{Power amplifier}
\acro{PCE}{Path Computation Element}
\acro{PCEP}{Path Computation Element Protocol}
\acro{PCF}{Path Computation Function}
\acro{PCF}{Point Coordination Function}
\acro{PDR}{Packet Delivery Ratio}
\acro{PDV}{Packet Delay Variation}
\acro{PER}{Packet Error Rate}
\acro{PLCP}{Physical Layer Convergence Protocol}
\acro{PLL}{Phase-Locked Loop}
\acro{PL}{path loss}
\acro{PLR}{Point of Local Repair}
\acro{PMIP}{Proxy Mobile IP}
\acro{PoE}{Power over Ethernet}
\acro{PPDU}{Physical Protocol Data Unit}
\acro{PPP}{Point-to-Point Protocol}
\acro{PSTN}{Public Switched Telephone Network}
\acro{PTP}{Precision Time Protocol}
\acro{PUE}{Power Usage Effectiveness}
\acro{PU}{Primary User}
\acro{QAM}{Quadrature amplitude modulation}
\acro{QoS}{Quality of Service}
\acro{RAN}{Radio Access Network}
\acro{RAND}{Random Channel Assignment}
\acro{RERR}{Route Error}
\acro{RFC}{Request for Comments}
\acro{RIPng}{Routing Information Protocol next generation}
\acro{RIP}{Routing Information Protocol}
\acro{RPC}{Remote Procedure Call}
\acro{RPP}{Received Packet Power}
\acro{RREP}{Route Reply}
\acro{RREQ}{Route Request}
\acro{RSSI}{Received Signal Strength Indication}
\acro{RSS}{Received Signal Strength}
\acro{RSVP}{Resource ReSerVation Protocol}
\acro{RSVP-TE}{Resource ReSerVation Protocol - Traffic Engineering}
\acro{RTK}{Real Time Kinematic}
\acro{RTP}{Real-time Transport Protocol}
\acro{RTS}{Ready-To-Send}
\acro{RTT}{Round Trip Time}
\acro{RRC}{Radio Resource Control}
\acro{SAA}{Stateless Address Autoconfiguration}
\acro{SAPOS}{Satellitenpositionierungsdienst der deutschen Landesvermessung}
\acro{SAP}{Service Access Point}
\acro{SBC}{Single-Board Computer}
\acro{SBM}{Subnetwork Bandwidth Manager}
\acro{SBR}{System zur Bestimmung des Richtungsfehlers}
\acro{SCADA}{Supervisory Control and Data Acquisition}
\acro{SC-FDMA}{Single Carrier Frequency Division Multiple Access}
\acro{SDMA}{Space-division multiple access}
\acro{SDN}{Software Defined Networking}
\acro{SDR}{Software Defined Radio}
\acro{SDWN}{Software Defined Wireless Networks}
\acro{SENF}{Simple and Extensible Network Framework}
\acro{SGI}{Short Guard Intervall}
\acro{SIFS}{Short Interframe Space}
\acro{SINR}{Signal-to-Noise-plus-Interference Ratio}
\acro{SIP}{Session Initiation Protocol}
\acro{SISO}{Single Input Single Output}
\acro{SNR}{signal-to-noise ratio}
\acro{SONET}{Synchronous Optical Networking}
\acro{SPoF}{Single Point of Failure}
\acro{SSID}{Service Set Identifier}
\acro{STP}{Spanning Tree Protocol}
\acro{SU}{Secondary User}
\acro{TCP}{Transmission Control Protocol}
\acro{TPC}{Transmission Power Control}
\acro{TC}{Topology Control}
\acro{TDMA}{Time Division Multiple Access}
\acro{TEEER}{Telecommunications Equipment Energy Efficiency Rating}
\acro{TEER}{Telecommunications Energy Efficiency Ratio}
\acro{TETRA}{Terrestrial Trunked Radio}
\acro{TE}{Traffic Engineering}
\acro{TIM}{Technology Independend Monitoring}
\acro{TLV}{Type-Length-Value}
\acro{TLS}{Transport Layer Security}
\acro{TORA}{Temporally-Ordered Routing Algorithm}
\acro{ToS}{Type of Service}
\acro{TSFT}{Time Synchronization Function Timer}
\acro{TTL}{Time to live}
\acro{TVWS}{TV White Space}
\acro{TXOP}{Transmit opportunity}
\acro{UAV}{Unmanned Aerial Vehicle}
\acro{UDLR}{Unidirectional Link Routing}
\acro{UDL}{Unidirectional Link}
\acro{UDP}{User Datagram Protocol}
\acro{UDT}{Unidirectional Technology}
\acro{UE}{User Equipment}
\acro{UHF}{Ultra High Frequency}
\acro{UMA}{Unlicensed Mobile Access}
\acro{UMTS}{Universal Mobile Telecommunications System}
\acro{U-NII}{Unlicensed National Information Infrastructure}
\acro{UPS}{Uninterruptible Power Supply}
\acro{UP}{User Priorities}
\acro{USB}{Univeral Serial Bus}
\acro{USO}{Universal Service Obligation}
\acro{USRP}{Universal Software Radio Peripheral}
\acro{VCO}{Voltage-Controlled Oscillator}
\acro{VoIP}{Voice-over-IP}
\acro{VPN}{Virtual Private Network}
\acro{WAN}{Wide Area Network}
\acro{WBN}{Coordinated Wireless Backhaul Network}
\acro{WDS}{Wireless Distribution System}
\acro{WiBACK}{Wireless Back-Haul}
\acro{Wi-Fi}{Wireless Fidelity}
\acro{WiLD}{WiFi-based Long Distance}
\acro{WiMAX}{Worldwide Interoperability for Microwave Access}
\acro{WISPA}{Wireless Internet Service Provider Assocication}
\acro{WISP}{Wireless Internet Service Provider}
\acro{WLAN}{Wireless Local Area Network}
\acro{WLC}{Wireless LAN Controller}
\acro{WMN}{Wireless Mesh Network}
\acro{WMN}{Wireless Mesh Network}
\acro{wmSDN}{Wireless Mesh Software Defined Network}
\acro{WNIC}{Wireless Network Interface Controller}
\acro{WN}{WiBACK Node}
\acro{WRAN}{Wireless Regional Area Network}
\acro{WSN}{Wireless Sensor Network}
\acro{ZigBee}{ZigBee Alliance IEEE 802.15.4}
\acro{ZPR}{Zone Routing Protocol}
\acro{ABP}{Activation by Personalization}
\acro{ADR}{Adaptive Data Rate}
\acro{IoT}{Internet of Things}
\acro{LPWAN}{low-power wide-area network}
\acro{LoRa}{Long Range}
\acro{LoRaWAN}{Long Range Wide Area Network}
\acro{GPRS}{General Packet Radio Service}
\acro{CEP}{Circular Error Probable}
\acro{EIRP}{equivalent isotropically radiated power}
\acro{ITM}{Longley-Rice Irregular Terrain Model}
\acro{SF}{Spreading Factor}
\acro{ASDU}{Application Service Data Unit}
\acro{IEC 104}{IEC 60870-5-104}
\acro{SIM}{Subscriber Identity Module}
\acro{eSIM}{embedded \ac{SIM}}
\acro{PKI}{Public Key Infrastructure}
\acro{FHSS}{Frequency-Hopping Spread Spectrum}
\acro{UWB}{Ultra-Wideband}
\acro{PUF}{Physical Unclonable Functions}
\acro{RSRP}{Reference Signal Received Power}
\acro{RSRQ}{Reference Signal Received Quality}
\acro{UI}{User Interface}
\acro{TAC}{Tracking Area Code}
\acro{CID}{Cell ID}
\acro{eNB}{eNodeB}
\acro{DRX}{Discontinuous Reception}
\acro{eDRX}{Extended Discontinuous Reception}
\acro{CMC}{Cipher Mode Command} %
\acro{MS}{Mobile Station}
\acro{BCCH}{Broadcast Common Control Channel}
\acro{AuC}{Authentication Centre}
\acro{FOSS}{Free and Open Source Software}
\acro{MS}{Mobile Station}
\acro{LFSR}{Linear-Feedback Shift Register}
\acro{TMTO}{Time-Memory Tradeoff}
\acro{LAC}{Location Area Code}
\acro{MNC}{Mobile Network Code}
\acro{IV}{Initialization Vector}
\acro{SI3}{System Information Message Type 3}
 
\end{acronym}